\documentclass[preprint,aps]{revtex4}

\usepackage{graphicx}

\begin{document}

\title{Persistence of Topological Order and Formation of Quantum Well States in Topological Insulators Bi$_2$(Se,Te)$_3$ under Ambient Conditions}
\author{Chaoyu Chen$^{1}$, Shaolong He$^{1}$, Hongming Weng$^{1}$, Wentao Zhang$^{1}$, Lin Zhao$^{1}$, Haiyun Liu$^{1}$, Xiaowen Jia$^{1}$, Daixiang Mou$^{1}$, Shanyu Liu$^{1}$, Junfeng He$^{1}$, Yingying Peng$^{1}$, Ya Feng$^{1}$, Zhuojin Xie$^{1}$, Guodong Liu$^{1}$, Xiaoli Dong$^{1}$, Jun Zhang$^{1}$, Xiaoyang Wang$^{2}$, Qinjun Peng$^{2}$, Zhimin Wang$^{2}$, Shenjin Zhang$^{2}$, Feng Yang$^{2}$, Chuangtian Chen$^{2}$, Zuyan Xu$^{2}$, Xi Dai$^{1}$, Zhong Fang$^{1}$, X. J. Zhou$^{1,*}$}

\affiliation{
\\$^{1}$Beijing National Laboratory for Condensed Matter Physics, Institute of Physics, Chinese Academy of Sciences, Beijing 100190, China
\\$^{2}$Technical Institute of Physics and Chemistry, Chinese Academy of Sciences, Beijing 100190, China}
\date{July 28, 2011}
%
% The abstract goes here
%

%%%\begin{abstract}

%%%%\end{abstract}

%%%%\pacs{74.70.-b, 74.25.Jb, 79.60.-i, 71.20.-b}

\maketitle

{\bf The topological insulators represent a unique state of matter where the bulk is insulating with an energy gap while the surface is metallic with a Dirac cone protected by the time reversal symmetry\cite{Tin3D,ZhangQiPT,HasanKaneRev,ZhangQiRMP,JMoorePer}. These characteristics provide a venue to explore novel quantum phenomena in fundamental physics\cite{XLQiNP,RDLiAxions,XLQiMomo,LFuMajorana,RYuSience} and show potential applications in spintronics and quantum computing\cite{Revnextgeneration,ZhangQiPT,JMoorePer}. One critical issue directly related with the applications as well as the fundamental studies is how the topological surface state will behave under ambient conditions (1 atmosphere air and room temperature). In this paper, we report high resolution angle-resolved photoemission measurements on the surface state of the prototypical topological insulators, Bi$_2$Se$_3$, Bi$_2$Te$_3$ and Bi$_2$(Se$_{0.4}$Te$_{2.6}$), upon exposing to ambient conditions.  We find that the topological order persists even when the surface is exposed to air at room temperature.  However, the surface state is strongly modified after such an exposure.  Particularly, we have observed the formation of two-dimensional quantum well states near the surface of the topological insulators after the exposure which depends sensitively on the original composition, {\it x},  in Bi$_2$(Se$_{3-x}$Te$_x$). These rich information are crucial in utilizing the surface state and in probing its physical properties under ambient conditions.
}

The angle-resolved photoemission spectroscopy (ARPES) is a powerful experimental tool to directly identify and characterize topological insulators\cite{HasanARPES}.  A number of three-dimensional topological insulators have been theoretically predicted and experimentally identified by ARPES\cite{BiSbARPES,BiSeTheo,Bi2Se3ARPES,Bi2Te3ARPES,TlBiSe2Theo,TlBiSe2ARPES}; some of their peculiar properties have been revealed by scanning tunneling microscopy (STM)\cite{TZhangSTM,YazdaniSTM,KaptulnikSTM,PCheng,Hanaguri}. The application of the topological surface states depends on the surface engineering that can be  manipulated by incorporation of non-magnetic\cite{LAWraypn,mnmi,ccvb,crs} or magnetic\cite{YLChenFe,ZHasanPertur,LAWraypn,mnmi} impurities or gas adsorptions\cite{ZHasanPertur,LPlucinski,LAWraypn,HMBenia}.  While the ARPES and STM measurements usually involve the fresh surface obtained by cleaving samples {\it in situ} under ultra-high vacuum,  for the transport measurements which are widely used to investigate the intrinsic quantum behaviors of the topological surface state\cite{DXQu,JGAnalytis,ZRen,JChen},  and particularly the ultimate utilizations of the topological insulators, the surface may be exposed to ambient conditions (1 atmosphere air and room temperature) or some gas protection environment. It is therefore crucial to investigate whether the topological order can survive under the ambient conditions, and furthermore, whether and how the surface state may be modified after such exposures.

We start by first looking at the electronic structure of the prototypical topological insulators Bi$_2$(Se,Te)$_3$ under ultra-high vacuum.  The Fermi surface and the band structure of the Bi$_2$(Se$_{3-x}$Te$_x$) topological insulators depend sensitively on the composition, x, as shown in Fig. 1.  The single crystal samples here were all  cleaved {\it in situ} and measured  at 30 K in an ultra-high vacuum (UHV) chamber with a base pressure better than 5 $\times$ 10$^{-11}$ Torr.   For Bi$_2$Se$_3$, a clear Dirac cone appears near -0.36 eV (Figs. 1d and 1e); the corresponding Fermi surface (Fig. 1a) is nearly circular but with a clear hexagon-shape in the measured data\cite{HexBi2Se3}. It is apparently of {\it n}-type because the Fermi level intersects with the bulk conduction band.  On the other hand, the Dirac cone of the Bi$_2$Te$_3$ sample lies near -0.08 eV (Figs. 1h and 1i), much closer to the Fermi level than that reported before (-0.34 eV in \cite{Bi2Te3ARPES}). The corresponding Fermi surface (Fig. 1c) becomes rather small, accompanied by the appearance of six petal-like bulk  Fermi surface sheets.  These results indicate that our Bi$_2$Te$_3$ sample is of {\it p}-type because the Fermi level intersects the bulk valence band along the $\overline{\Gamma}$-$\overline{M}$ direction. This is also consistent with the positive Hall coefficient measured on the same Bi$_2$Te$_3$ sample\cite{CZhang}.  This difference of the Fermi surface topology and the location of the Dirac cone from others\cite{Bi2Te3ARPES}  may be attributed to the different carrier concentration in Bi$_2$Te$_3$ due to different sample preparation conditions. In our Bi$_2$(Se$_{3-x}$Te$_x$) samples, we have seen a crossover from n-type Bi$_2$Se$_3$  to p-type Bi$_2$Te$_3$.   In order to eliminate the interference of the bulk bands on the  surface state near the Fermi level, we fine tuned the composition {\it x} in Bi$_2$(Se$_{3-x}$Te$_x$) and found that, for {\it x}=2.6,  nearly no spectral weight can be discerned from the bulk conduction band, as seen from both the Fermi surface (Fig. 1b) and the band structure (Figs. 1f and 1g). A slight substitution of Te by Se in Bi$_2$(Se$_{0.4}$Te$_{2.6}$) causes a dramatic drop of the Dirac point to -0.31 eV (Figs. 1f and 1g) and an obvious hexagon-shaped Fermi surface (Fig. 1b). It is interesting to note that the hexagon-shape of Bi$_2$(Se$_{0.4}$Te$_{2.6}$) (Fig. 1b)  is rather pronounced, although its Fermi surface size is smaller than that of Bi$_2$Se$_3$ (Fig. 1a).  The hexagonally-shaped Fermi surface observed in the topological surface states reflects the hybridization of surface electronic states with the bulk states and can be theoretically explained by considering the higher order terms in the $\emph{k}$ $\cdot$ $\emph{p}$ Hamiltonian\cite{HexWarp}.

In order to directly examine how the topological surface state behaves under ambient conditions in the topological insulators, we carried out our ARPES measurements in different ways. (1). We first cleaved the sample \emph{in situ} and performed ARPES measurement  in the ultra-high vacuum (UHV) chamber.  The sample was then pulled out to another chamber filled with 1 atmosphere N$_2$ gas,  exposed for about 5 minutes, before transferring back to the UHV chamber to do ARPES measurements;  (2). We cleaved and measured the sample in the UHV chamber,  and then pulled the sample out to air for 5 minutes before transferring back to the UHV chamber for the ARPES measurements;  (3). We cleaved the sample in air and then transferred it to the UHV chamber to do the ARPES measurement.  Our measurements show that the above procedures of exposure to air or N$_2$  produce similar and reproducible results for a given sample.

The surface exposure of the topological insulators to air or N$_2$ gives rise to a dramatic alteration of the surface state, as shown in Figs. 2, 3 and 4, for Bi$_2$Se$_3$, Bi$_{2}$(Se$_{0.4}$Te$_{2.6}$), and Bi$_2$Te$_3$, respectively, when compared with those for the fresh surface (Fig. 1).  The first obvious change is the shifting of the Dirac cone position relative to the Fermi level. For Bi$_2$Se$_3$, Bi$_{2}$(Se$_{0.4}$Te$_{2.6}$) and  Bi$_2$Te$_3$, it shifts from the original -0.36 eV (Figs. 1d and 1e), -0.31 eV (Figs. 1f and 1g), -0.08 eV (Figs. 1h and 1i) for the fresh surface to -0.48 eV (Fig. 2b), -0.40 eV (Figs. 3a and 3b), and  -0.28 eV [Figs. 4(c-f)] at 30 K for the exposed surface, respectively. In all these cases,  the shift of the Dirac cone to a larger binding energy indicates an additional doping of electrons into the surface state.   The exposure also gives rise to a dramatic change of the surface Fermi surface.  For Bi$_2$Se$_3$, in addition to a slight Fermi surface size increase, an obvious change occurs in the Fermi surface shape that the hexagon-shape becomes much more pronounced in the exposed surface (Fig. 2d) than that in the fresh sample (Fig. 1a). For Bi$_{2}$(Se$_{0.4}$Te$_{2.6}$), one clearly observes the much-enhanced warping effect in the exposed surface (Fig. 3c) when compared with the nearly standard hexagon in the fresh surface (Fig. 1b).  The most dramatic change occurs for Bi$_2$Te$_3$ where not only the Fermi surface size increases significantly, but also the warping effect in the exposed surface (Fig. 4i) becomes much stronger. Overall, the exposure causes the lowering of the Dirac cone position,  an increase of the surface Fermi surface size, and an obvious enhancement of the Fermi surface warping effect in the Bi$_2$(Se$_{3-x}$Te$_x$) system.

The topological order in the Bi$_2$(Se$_{3-x}$Te$_x$) topological insulators is robust even when the surface is exposed to ambient conditions, in spite of all the alterations mentioned above.  One clearly observes the persistence of the Dirac cone in the exposed surface as in Bi$_2$Se$_3$ (Figs. 2b and 2c), in Bi$_{2}$(Se$_{0.4}$Te$_{2.6}$) (Figs. 3a and 3b), and Bi$_2$Te$_3$ (Figs. 4(c-h)). Particularly, this is the case for the surface exposed to air and measured at room temperature (Fig. 2c for Bi$_2$Se$_3$, and Figs. 4g and 4h for Bi$_2$Te$_3$).  On the other hand, after the exposure, although the signal of the surface state gets weaker for Bi$_2$Se$_3$ (Fig. 2), it remains rather strong for Bi$_{2}$(Se$_{0.4}$Te$_{2.6}$) (Fig. 3) and Bi$_2$Te$_3$ (Fig. 4). This is in a stark contrast to the conventional trivial surface state where minor surface contamination will cause the extinction of the surface state\cite{Huefner}. The robustness of the topological order to Coulomb, magnetic and disorder perturbations has been reported before\cite{ZHasanPertur,LPlucinski}. Our present observations directly demonstrate the robustness of the topological order against absorption and thermal process under ambient conditions due to the protection of the time-reversal symmetry\cite{HasanKaneRev,ZhangQiRMP}.

The surface exposure to air or N$_2$ in the Bi$_2$(Se$_{3-x}$Te$_x$) topological insulators produces two-dimensional electronic states near the surface.  In Bi$_2$Se$_3$, the exposure gives rise to additional parabolic bands,  as schematically marked by the dashed line in Figs. 2b and 2c. Correspondingly, this leads to additional Fermi surface sheet(s) inside the regular topological surface state (Figs. 2d and 2e).  In Bi$_{2}$(Se$_{0.4}$Te$_{2.6}$), this effect gets more pronounced and the newly emerged bulk conduction band splits into several discrete bands, as marked by the dashed lines in Fig. 3b.   While the band quantization effect occurs in the bulk conduction band in Bi$_{2}$(Se$_{0.4}$Te$_{2.6}$), it shows up in the valence band in the exposed Bi$_2$Te$_3$ surface, as shown in Figs. 4(c-h), where one can see a couple of discrete ``M"-shaped bands. The quantized bands are obvious at low temperature and get slightly smeared out when the temperature rises to room temperature (Figs. 4g and 4h).

The formation of the split bands in the exposed surface of the topological insulators is reminiscent of the quantum well states observed in the quantum confined systems\cite{SiAg} and in some topological insulators\cite{tdeg, mnmi, crs, ccvb,HMBenia,KingReshba}. There are a couple of possibilities that the quantum well states may be formed.  One usual way is due to band bending effect. As mentioned before, the surface exposure to air or N$_2$ causes an electron transfer to the surface of the topological insulators.  The accumulation of these additional electrons near the surface would lead to a downward bending of the bulk bands near the surface region, as schematically shown in Fig. 3d, resulting in a ``V"-shaped potential well where the bulk conduction band of electrons can be confined. This picture, as proposed before\cite{tdeg,ccvb}, seems to be able to explain the two-dimensional quantum well states in the conduction bands in Bi$_{2}$(Se$_{0.4}$Te$_{2.6}$)  (Figs. 3a and 3b).  However, it becomes questionable to explain the quantum well states observed in the bulk valence band of Bi$_2$Te$_3$ (Figs. 4c-j). In this case, the downward band bending no longer acts as a quantum well potential for the valence band top because the charge carriers are hole-like. The band-bending is therefore not a general picture that can explain the formation of the two-dimensional quantum well states in Bi$_{2}$(Se$_{0.4}$Te$_{2.6}$)  (Figs. 3a and 3b) and Bi$_2$Te$_3$ (Fig. 4) topological insulators on the same footing.   An alternative scenario is the expansion of van der Waals spacings in between the quintuple layers (QLs) caused by the intercalation of gases\cite{SVEremeev}.  The observation of multiple split bands with different spacings would ask for multiple van der Waals gaps with different expansions.  Whether and how these can be realized in the exposed surface remains to be investigated.  We note that our observations of multiple split bands are similar to those seen in the ultra-thin films of Bi$_2$Se$_3$\cite{YZhang} and Bi$_2$Te$_3$\cite{YYLi}.  From our first principle band structure calculations on Bi$_2$Te$_3$ with different number of quintuple layers,  we also find that a detached slab with a thickness of 7 quintuple layers can give a rather consistent description (Fig. 4l) of our observed results in terms of the quantitative spacings between the 3 resolved bands (VB0, VB1, and VB2 bands as marked in Figs. 4c and 4l). In addition, the distance between the conduction band bottom (CB0 band in Figs. 4i and 4l) and the first valence sub-band bottom (VB0 band in Figs. 4i and 4l) is rather consistent between the measured and calculated results.  These seem to suggest that a  ``confined surface slab" with nearly 7 quintuple layers may be formed after the exposure that acts more or less independently from the bulk.  More work needs to be done to further investigate whether such a confined surface slab can be thermodynamically stable.  Overall, the formation of the two-dimensional quantum well states is a general phenomenon for the exposed surface of the Bi$_2$(Se$_{3-x}$Te$_x$) topological insulators; the effect depends sensitively on the composition {\it x} of the samples which may facilitate manipulation of these quantum well states.

The present work has significant implications on the fundamental study and ultimate applications of the topological insulators. Many experimental measurements, such as some transport measurements,  involve samples exposed to ambient conditions. The practical applications may involve sample surface either exposed to ambient condition, or be in contact with other magnetic or superconducting materials.  On the one hand, the robustness of the topological order under ambient conditions sends a good signal for these experimental characterization and practical utilizations.  The formation of the quantum well states may give rise to new phenomena to be studied and utilized. The sensitivity of the surface state to the Bi$_2$(Se$_{3-x}$Te$_x$) composition provides a handle to manipulate these quantum states.  On the other hand, the strong modification of the electronic structure and the formation of additional quantum well states in the exposed surface have to be considered seriously in interpreting experimental data and in surface engineering.   It is critical to realize before hand that the surface under study or to be utilized may exhibit totally different behaviors as those from the fresh surface cleaved in ultra-high vacuum.  In addition to the alteration of electronic states upon exposure, the transport properties of the topological surface state may be further complicated by the formation of quantum well states.

We thank Prof. X. H. Chen for providing us samples at the initial stage of the project, and Prof. Liling Sun and Prof. Zhong-xian Zhao for their help in the characterization of the samples. This work is supported by the NSFC (91021006) and the MOST of China (973 program No: 2011CB921703).

%%Experimental Description

{\bf\textbf{ METHODS} }

{\bf Crystal growth methods }
Single crystals of $Bi_{2}(Se_{3-x}Te_{x})$ (x=0, 2.6 and 3) were grown by the self-flux method. Bismuth, selenium and tellurium powders were weighed according to the stoichiometric $Bi_{2}(Se_{3-x}Te_{x})$ (x=0, 2.6 and 3) composition. After mixing thoroughly, the powder was placed in alumina crucibles and sealed in a quartz tube under vacuum. The materials were heated to 1000 $^\circ$C, held for 12 hours to obtain a high degree of mixing, and then slowly cooled down to 500 $^\circ$C over 100 hours before cooling to room temperature. Single crystals of several millimeters in size were obtained. The crystal structure of the resulting crystals was examined by use of a rotating anode x-ray diffractometer with Cu \emph{K$_{\alpha}$} radiation ($\lambda$ = 1.5418 {\AA}). The chemical composition of the crystals was analyzed by the energy dispersive X-ray spectroscopy (EDAX) and the induction-coupled plasma atomic emission spectroscopy (ICP-AES).  The resistivity of the crystals was measured by the standard four-probe method.

{\bf Laser-ARPES methods. }
The angle-resolved photoemission measurements were carried out on our vacuum ultra-violet (VUV) laser-based angle-resolved photoemission
system\cite{LiuARPES}. The photon energy of the laser is 6.994 eV with a bandwidth of 0.26 meV. The energy resolution of the electron energy
analyzer (Scienta R4000) is set at 1 meV, giving rise to an overall energy resolution of $\sim$1 meV which is significantly improved
from 10$\sim$15 meV  from regular synchrotron radiation systems\cite{Bi2Se3ARPES,Bi2Te3ARPES}. The angular resolution is $\sim$0.3$^\circ$,
corresponding to a momentum resolution  $\sim$0.004 $\AA$$^{-1}$ at the photon energy of 6.994 eV, more than twice improved from 0.009 $\AA$$^{-1}$ at a regular photon energy of 21.2 eV for the same angular resolution. Our superior instrumental resolution of laser ARPES has made the measured features of topological insulators in this work much sharper.  The Fermi level is referenced by measuring on a clean polycrystalline gold that is electrically connected to the sample. The samples were all measured in vacuum with a base pressure better than 5$\times$10$^{-11}$ Torr.

$^{*}$Corresponding author: XJZhou@aphy.iphy.ac.cn

\newpage

%%\begin{figure}[tbp]
\begin{figure*}[tbp]
\begin{center}
\includegraphics[width=1.0\columnwidth,angle=0]{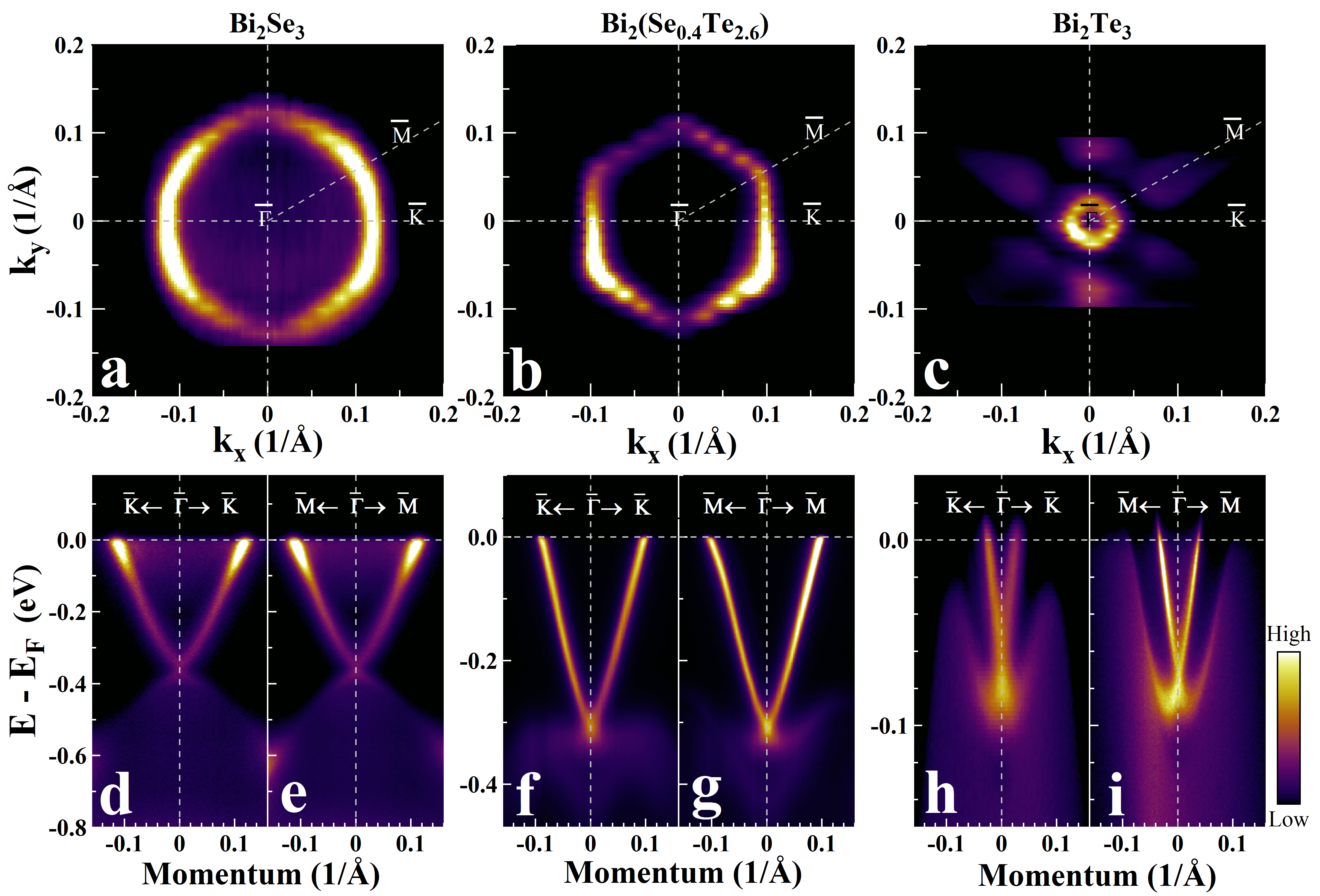}
\end{center}
\caption{Fermi surface and band structure of Bi$_{2}$(Se$_{3-x}$Te$_{x}$) (x=0, 2.6, 3) topological insulators cleaved {\it in situ} and measured at 30 K in ultra-high vacuum. (a-c) show Fermi surface of Bi$_{2}$Se$_{3}$, Bi$_{2}$(Se$_{0.4}$Te$_{2.6}$) and Bi$_{2}$Te$_{3}$, respectively. The Fermi surface here, and in other figures below, are original data without involving artificial symmetrization.  The band structures along two high symmetry lines $\bar{\Gamma}-\bar{K}$ and $\bar{\Gamma}-\bar{M}$ are shown in (d,e) for Bi$_{2}$Se$_{3}$,  in (f,g) for Bi$_{2}$(Se$_{0.4}$Te$_{2.6}$) and (h,i) for Bi$_{2}$Te$_{3}$.
}
\end{figure*}
%%\end{figure}

%%\begin{figure}[tbp]
\begin{figure*}[tbp]
\begin{center}
\includegraphics[width=1.0\columnwidth,angle=0]{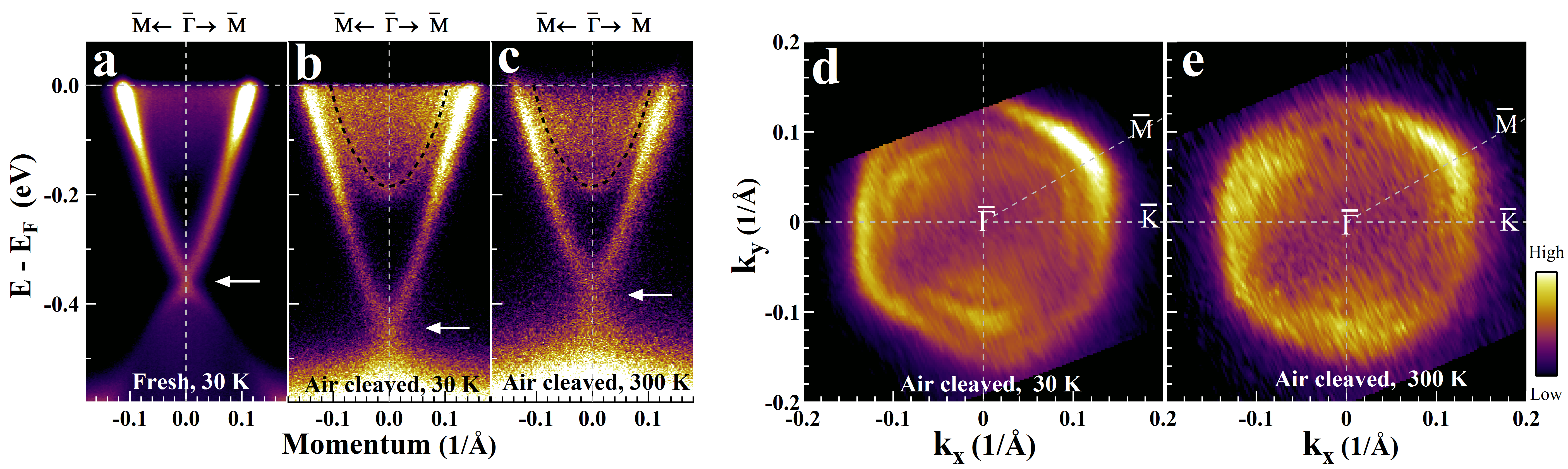}
\end{center}
\caption{Fermi surface and band structure of Bi$_{2}$Se$_{3}$ cleaved in air and measured in the ultra-high vacuum (UHV) chamber. (a). Band structure of the fresh Bi$_{2}$Se$_{3}$ cleaved and measured in the UHV chamber at 30 K along  $\bar{\Gamma}-\bar{M}$ direction.   (b). Band structure of Bi$_{2}$Se$_{3}$ cleaved in air and measured in UHV at 30 K along  $\bar{\Gamma}-\bar{M}$ direction.  (c). Band structure of Bi$_{2}$Se$_{3}$ cleaved in air and measured in UHV at 300 K along  $\bar{\Gamma}-\bar{M}$ direction. (d,e). Fermi surface of Bi$_{2}$Se$_{3}$ cleaved in air and measured in UHV at 30K and 300 K, respectively.   Black dashed lines in (b) and (c) mark the parabolic bands above the Dirac point from the two-dimensional electron gas.}
\end{figure*}
%%\end{figure}

%%\begin{figure}[b]
\begin{figure*}[tbp]
\begin{center}
\includegraphics[width=1.0\columnwidth,angle=0]{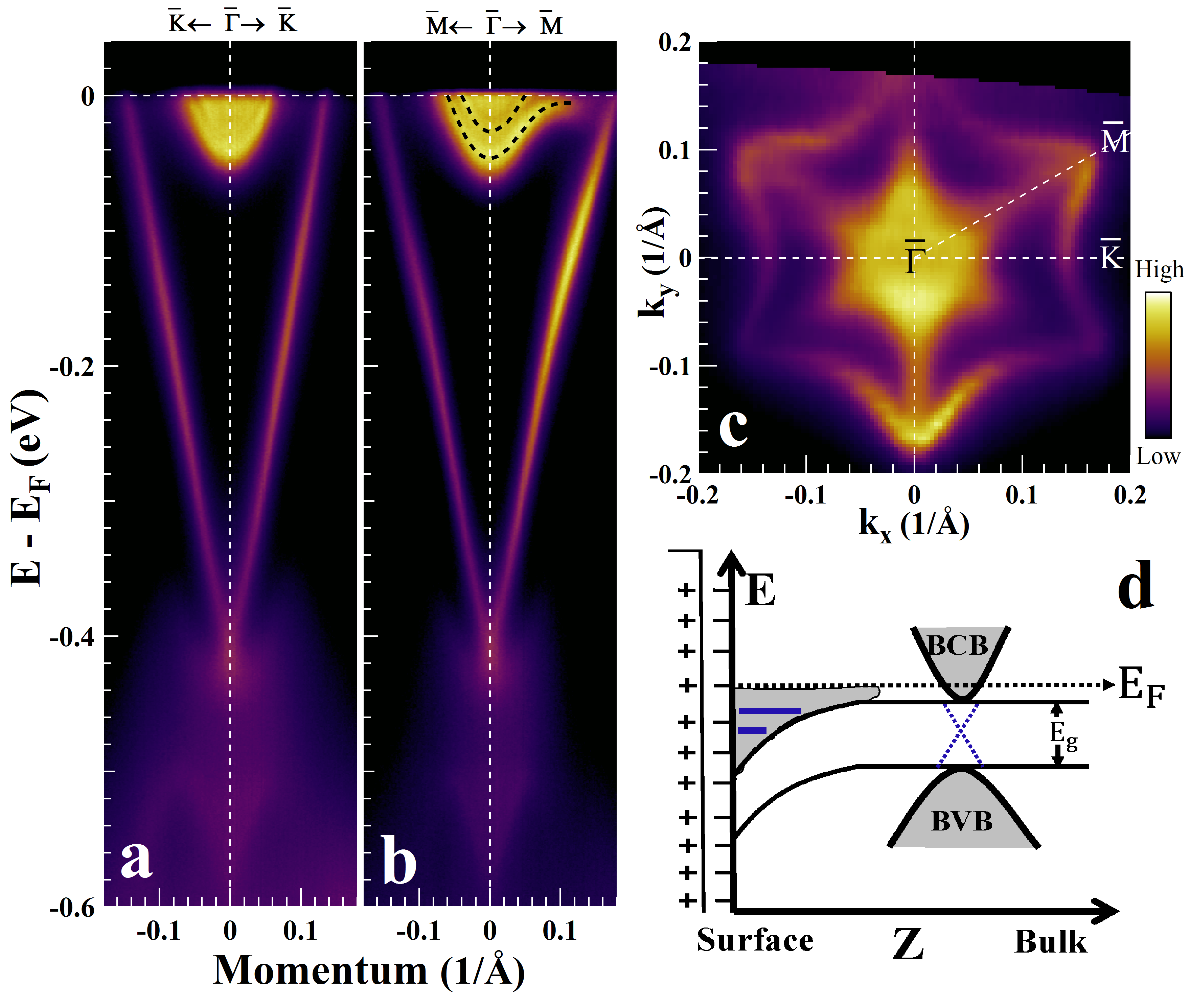}
\end{center}
\caption{Emergence of quantum well states in Bi$_{2}$Te$_{2.6}$Se$_{0.4}$ after exposing to N$_{2}$. (a,b). Band structure  measured at 30 K along $\bar{\Gamma}-\bar{K}$ and $\bar{\Gamma}-\bar{M}$, respectively. Black dashed lines in (b) mark the quantum well states formed in the bulk conduction band (BCB) above the Dirac point. (c). The corresponding Fermi surface. It shows three-fold symmetry where three corners of M points are strong while the other three are weak. This is also in agreement with the asymmetric band structure in Fig. 3b.  (d). Schematic band structure showing the possible formation of the quantum well states near the sample surface in the bulk conduction band. The blue dotted lines between the bulk valence band (BVB) and bulk conduction band (BCB) represent the topological surface states while the blue solid lines represent quantum well states.
}
\end{figure*}
%%\end{figure}

%%\begin{figure}[t]
\begin{figure*}[tbp]
\begin{center}
\includegraphics[width=1.0\columnwidth,angle=0]{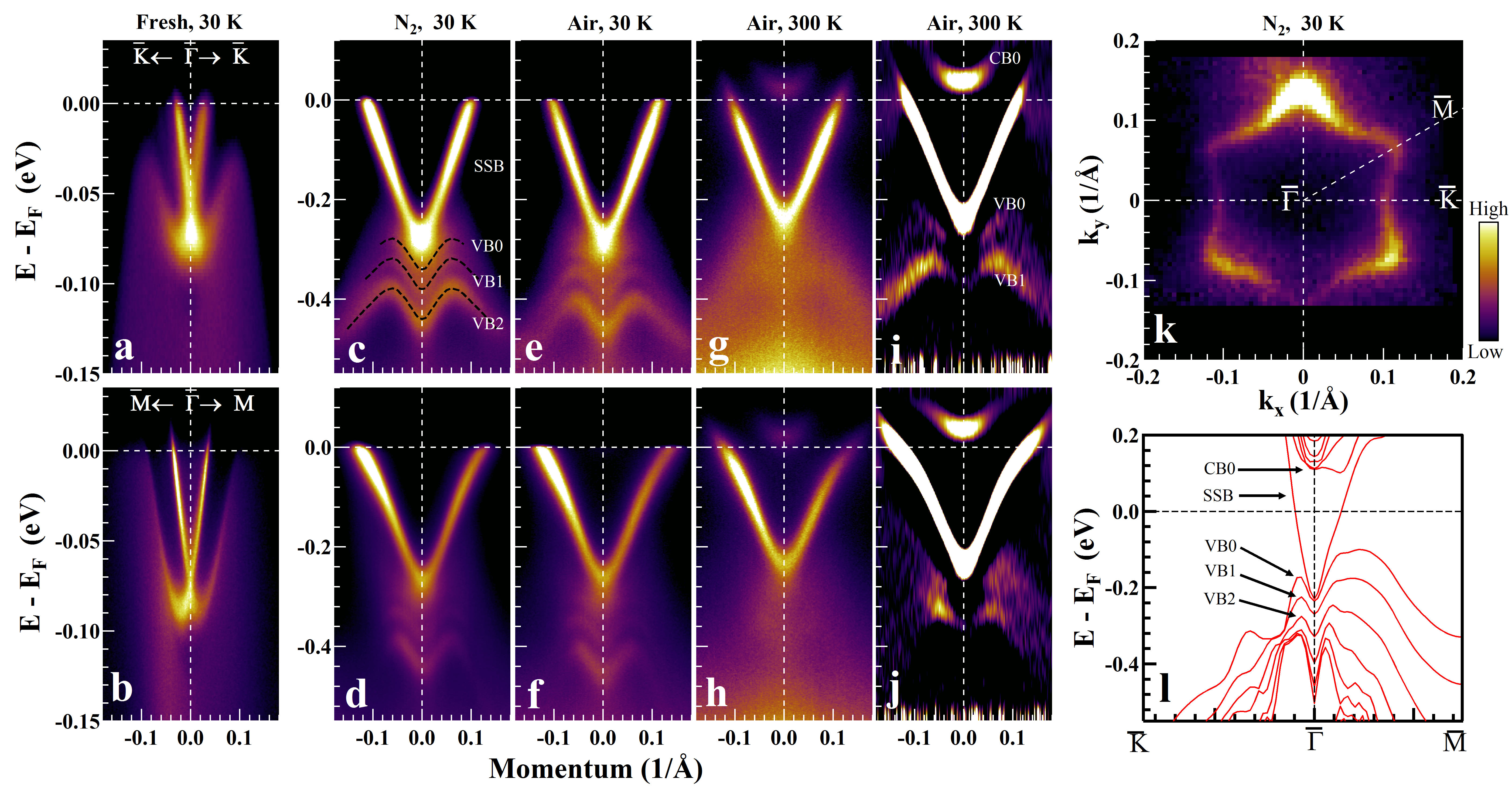}
\end{center}
\caption{Persistence of topological surface state and formation of quantum well states in Bi$_{2}$Te$_{3}$ after exposure to N$_2$ or air. The sample was first cleaved and measured in UHV at 30 K. (a,b) show the corresponding band structure along the $\bar{\Gamma}-\bar{K}$ and $\bar{\Gamma}-\bar{M}$ directions. The sample was then pulled out from the UHV chamber and exposed to N$_2$ at 1 atmosphere for 5 minutes before transferring back into UHV chamber for the ARPES measurement. (c,d) show the band structure of the N$_2$-exposed sample along the $\bar{\Gamma}-\bar{K}$ and $\bar{\Gamma}-\bar{M}$ directions. The black dashed lines in (c) illustrate the quantum well  states formed in the bulk valence band below the Dirac point.  The sample was then pulled out again and exposed to air for 5 minutes before putting back in vacuum for ARPES measurement. (e,f) show the band structure of the air-exposed sample at 30 K along the $\bar{\Gamma}-\bar{K}$ and $\bar{\Gamma}-\bar{M}$ directions.  (g,h) show the measurements at 300 K and (i,j) show their corresponding second-derivative images in order to highlight the bands. (k) Fermi surface of N$_2$-exposed sample. (l). First principle calculation of the band structure of Bi$_2$Te$_3$ slab with seven quintuple layers.
}
\end{figure*}
%%\end{figure}

\end{document}